\documentclass[12pt]{article}
\usepackage{amsmath}
\usepackage{epsfig}
\begin{document}
\title{On the loop approximation in nucleon QCD sum rules}
\author{E. G. Drukarev, M. G. Ryskin, V. A. Sadovnikova\\
{\em National Research Center "Kurchatov Institute"}\\
{\em B. P. Konstantinov Petersburg Nuclear Physics Institute}
}
\date{\today}
\maketitle

\begin{abstract}
There was a general believe that the nucleon QCD sum rules which include only the quark loops and thus contain only the condensates of dimension $d=3$ and $d=4$ have only a trivial solution. We demonstrate that there is also a
nontrivial solution. We show that it can be treated as the lowest order approximation to the solution
which includes the higher terms of the Operator Product Expansion. Inclusion of the radiative corrections improves the convergence of the series.
\end{abstract}

\section{Introduction}

The QCD sum rules (SR) approach suggested in \cite{1} for description of the static properties of mesons and later employed for nucleons \cite{2,3} is based on the dispersion relation for the
 function describing the propagation of the system which carries the quantum
 numbers of the hadron. In the case of the proton this function, usually referred to as the "polarization operator",
 can be written as
\begin{equation}
\Pi(q)=\hat q\Pi_q(q^2)+I\Pi_i(q^2),
\label{1}
\end{equation}
with $q$ the four-momentum of the system, $\hat q=q_{\mu}\gamma^{\mu}$, $I$ is the unit matrix. The dispersion relations
 \begin{equation}
\Pi_i(q^2)=\frac1\pi\int dk^2\frac{\mbox{Im}\Pi_i(k^2)}{k^2-q^2}\,;
 \quad i=q,I.
\label{2}
\end{equation}
are considered at $q^2\to-\infty$. This enables to expand the left hand sides (LHS) of Eq.~(\ref{2}) in powers of $1/q^2$. The coefficients of the expansion are the QCD condensates.This is known as the Operator Product Expansion (OPE) \cite{3}.
This provides the perturbative expansion of the short
distance effects, while the nonperturbative physics is contained in the
condensates. The higher terms of the OPE contain the condensates of the higher dimension.

The left hand side of Eq.~(\ref{2}) can be written as
\begin{equation}
\Pi_q^{OPE}(q^2)=\sum_{n}A_n(q^2); \quad \Pi_I^{OPE}(q^2)=\sum_{n} B_n(q^2)
\label{3a}
\end{equation}
where the lower index $n$ is the dimension of the corresponding QCD condensate. The lowest order OPE contribution to the structure $\Pi_q^{OPE}(q^2)$ is the quark loops, and does not contain condensates. We denote it as $A_0$.
The lowest order OPE contribution to the structure $\Pi^{I~OPE}(q^2)$ is proportional to the scalar quark condensate $\langle 0|\bar q q|0\rangle$, which has dimension $d=3$. This contribution contains a quark loop. Next comes the term with the gluon condensate $\langle 0|\frac{\alpha_s}{\pi}G^{a\mu \nu}G^a_{\mu \nu}|0\rangle$ with dimension $d=4$, also containing the quark loops.
The OPE terms of the higher dimension do not contain loops.
The loops provide the logarithmic terms $(q^2)^n\ln {q^2}$ on the left hand side, while the contributions with $d>4$ provide the power corrections $(q^2)^{-n}$. The physical spectrum of polarization operator contains the single-particle states $p, N^*$, etc. and the many-particle states, e.g. protons and pions. The former manifest themselves as the poles of the integrand on the right hand side (RHS) of Eq.~(\ref{2}) while the latter correspond to its cuts.
Both the logarithmic terms and the power corrections on the LHS contribute to the terms originated by the poles and the cuts on the RHS.

The RHS of Eq.~(\ref{2}) is usually approximated by the "pole+continuum" model
\cite{1,2} in which the lowest lying pole is written down exactly,
while the higher states are described by continuum. The spectral function on the cut corresponding to continuum is approximated by discontinuity of the logarithmic terms on the LHS.
Under the latest assumption Eqs.~(\ref{2}) take the form
\begin{equation}
\Pi_i^{OPE}(q^2)=\frac{\lambda_N^2\xi_ii}{m^2-q^2}+\frac1{2\pi i}
\int\limits_{W^2}^\infty dk^2\frac{\Delta\Pi_i^{OPE}(k^2)}{k^2-q^2};
\quad i=q,I.  \label{3}
\end{equation}
Here $\xi_q=1$,\ $\xi_I=m$. The upper index OPE means that several lowest OPE terms are included.
One usually applies the Borel transform [1-3] to both sides of Eq.~(\ref{2}), which converts the functions of $q^2$ to the functions
of the Borel mass $M^2$. An important assumption is that there
is an interval of the values of $M^2$ where the two sides of the SR have a good
overlap, approximating also the true functions. The standard interval ("duality interval") is $0.8$ GeV$^2 \leq M^2 \leq 1.4 $ GeV$^2$.

The proton mass $m$, the residue at the nucleon pole $\lambda^2$ and the continuum threshold $W^2$ are the unknowns of the QCD SR equations. The set of parameters $m,\lambda^2, W^2$ which minimize the function
\begin{equation}
\chi_N^2(m,\lambda^2, W^2)=\frac{1}{2N\sigma^2
}\sum_{j=1}^{j=N}\sum_{i=q,I}\Big(\frac{{\cal L}_i(M_j^2)-{\cal R}_i(M_j^2)}
{{\cal L}_i(M_j^2)}\Big)^2,
\label{3x}
\end{equation}
with ${\cal L}_i$ and ${\cal R}_i$ the Borel transforms of the RHS and LHS of Eq.(\ref{3}) correspondingly is referred to as the solution of the QCD sum rules equations. Here $M_j$ is a set of $N$ points in the duality interval. We put $\sigma=0.1$.

The "pole +continuum" model for the spectrum makes sense only if the contribution of the proton pole (treated exactly) to the RHS of Eq.~(\ref{3}) exceeds that of the continuum (treated approximately). We treat a solution which satisfy this condition as the physical one. Otherwise we call it an unphysical solution \cite{8}.

In order to obtain the value of $m$ close to the physical value $m \approx 940 $ MeV one should include the OPE terms up to those with $d=9$.
If only the loop terms $A_0$, $B_3$ and $A_4$ are included on the LHS of Eq.~(\ref{3}) ("loop approximation"), we immediately find a trivial solution $\lambda^2=0$, $W^2=0$, in which the logarithmic terms on the LHS correspond to the second term on the RHS of Eq.(\ref{3}). This solution exists for any value of the nucleon mass $m$. There was a general believe that this is
the only solution  in the loop approximation. This was viewed as one of the week points of the "pole+continuum" model and of the QCD SR approach as a whole \cite{4}.

Here we demonstrate that limiting ourselves to
the terms with $d \leq 4$ and even with $d \leq 3$ we have also another solution, which can be treated as a good lowest order approximation to the solution with $d \leq 9$.

\section{Equations in the loop approximation}

Keeping only the terms which contain the logarithmic loops we find
$\Pi_q^{OPE}(q^2)=A_0(q^2)+A_4(q^2)$ and $\Pi_I^{OPE}(q^2)=B_3(q^2)$.
The explicit form of the contributions depends on the form of the local operator $j(x)$ with the proton quantum numbers,
often referred to as ``current". The form of $j(x)$ is not unique, and we assume $j(x)=(u^T_a(x)C\gamma_{\mu}u_b(x))\gamma_5 \gamma^{\mu}d_c(x) \varepsilon^{abc}$, suggested in \cite{5}.
One can find \cite{2}, \cite{3}
\begin{equation}
A_0=\frac{-Q^4\ln{(Q^2/C^2)}}{64\pi^4}\,r_0; \quad B_3=\frac{aQ^2\ln{(Q^2/C^2)}}{16\pi^4}\,r_3 ;\quad
A_4=\frac{-b\ln({Q^2}/C^2)}{128\pi^4}\,r_4,
\label{3c}
\end{equation}
with $C^2$ the ultraviolet cutoff, $Q^2=-q^2>0$, while $a$ and $b$ are the scalar quark condensate and the gluon condensates multiplied by certain numerical factors
\begin{equation}
a=-(2\pi)^2\langle 0|\bar q q|0\rangle  ; \quad b=(2\pi)^2\langle 0|\frac{\alpha_s}{\pi}G^{a\mu \nu}G^a_{\mu \nu}|0\rangle.
\label{3d}
\end{equation}
The factors $r_d$ include the radiative corrections. Following \cite{6} we consider several cases.\

I). The radiative corrections are neglected, in this case $r_0=r_3=r_4=1$.\

II). Leading logarithmic approximation (LLA), in which only the terms $(\alpha_s\ln{Q^2})^n$ are included, but they are summed to all orders of $n$.\

III). The LLA terms and the nonlogarithmic corrections of the order $\alpha_s$ are taken into account.\

IV). All radiative corrections are included in the lowest order of $\alpha_s$.\

The LLA terms are expressed in terms of the function
\begin{equation}
L(Q^2)=\Big(\frac{\ln Q^2/\Lambda^2}{\ln \mu^2/\Lambda^2}\Big)^{4/9}.
\label{4a} \end{equation}
In Eq.~(\ref{4a}) $\Lambda=\Lambda_{QCD}$ is the QCD scale, while $\mu$ is the
normalization point, the standard choice is $\mu=0.5\,$GeV.
In the case II we have $r_0=r_4=L^{-1}$, while $r_3=1$. In the case III \cite{7}, \cite{6}
\begin{equation}
r_0=(1+\frac{71}{12}\frac{\alpha_s}{\pi})L^{-1}; \quad r_3=1+\frac{3}{2}\frac{\alpha_s}{\pi},
\label{4}
\end{equation}
with $\alpha_s=\alpha_s(1~$GeV$^2)$.
In the case IV the value of $r_0$ changes to
\begin{equation}
r_0=1+\frac{71}{12}\frac{\alpha_s}{\pi}- \frac{1}{2}\frac{\alpha_s}{\pi}\ln{\frac{Q^2}{\mu^2}},
\label{5}
\end{equation}
while $r_3$ remains the same as in the case II.
Since the contribution of the term $A_4$ is numerically small, we do not include the nonlogarithmic
corrections here, keeping $r_4=1/L$ for the cases II and III.

Actually the standard procedure is to consider the SR for the operators $\Pi^{OPE}_{i1}(q^2)=32\pi^4\Pi_i^{OPE}(q^2)$.
The factor $32\pi^4$ is
introduced in order to deal with the values of the order of unity (in
GeV units).

Finally the SR equations in the loop approximation, except the case IV can be written as
\begin{equation}
M^6E_2(\gamma)r_0(M^2,\gamma)+\frac{bM^2E_0(\gamma)}{4}r_4(M^2)=\lambda^2e^{-m^2/M^2},
\label{22}
\end{equation}
and
\begin{equation}
2aM^4E_1(\gamma)r_3=m\lambda^2e^{-m^2/M^2},
\label{22a}
\end{equation}
with $\gamma=W^2/M^2$, while
\begin{equation}
E_0(\gamma)=1-e^{-\gamma}, \quad E_1(\gamma)=1-(1+\gamma)e^{-\gamma}, \quad E_2(\gamma)=1-(1+\gamma+\gamma^2/2)e^{-\gamma}.
\label{23} \end{equation}
If all radiative corrections are neglected (I), $r_0=r_3=r_4=1$. In the LLA (II) $r_0=r_4=1/L(M^2)$, while $r_3=1$.
In these cases Eq.~(\ref{22}) and Eq.(\ref{22a}) are just the lowest terms of the standard nucleon sum rules [2,3].
Adding the  nonlogarithmic corrections of the order $\alpha_s$ to the LLA terms we obtain (III)
\begin{equation}
r_0=(1+\frac{71}{12}\frac{\alpha_s}{\pi})L^{-1}(M^2); \quad r_3=1+\frac{3}{2}\frac{\alpha_s}{\pi}.
\label{24}
\end{equation}
For the case when all corrections of the order $\alpha_s$ are treated perturbatively (IV), the first term on the LHS of Eq.~(\ref{22}) takes a more complicated form. It should be modified as \cite{6}
\begin{equation}
M^6E_2r_0 \rightarrow M^6E_2\Big[1+\frac{\alpha_s}{\pi}(\frac{53}{12}-\ln{\frac{W^2}{\mu^2}})\Big]-\frac{\alpha_s}{\pi}
\Big[M^4W^2(1+\frac{3}{4}\gamma)e^{-\gamma}+M^6{\cal E}(-\gamma)\Big],
\label{25}
\end{equation}
with ${\cal E}(x)=\sum_{n=1}x^n/(n\cdot n!)$.

\section{Solution of the equations}

Now we try to find the values of $m$, $\lambda^2$ and $W^2$ which minimize the difference between the LHS and RHS of Eqs.~(\ref{22}) and (\ref{22a}). We employ the value $\Lambda^2=230 $ MeV. Thus in the radiative corrections $\alpha_s(1 $GeV$^2)=0.475$. We vary the value of the scalar condensate around the point $\langle 0|\bar q q|0\rangle=(-241 $MeV$)^3$, corresponding to $a=0.55 $ GeV$^3$
We present also the results for $a=0.50 $ GeV$^3$ corresponding to $\langle 0|\bar q q|0\rangle=(-233 $MeV$)^3$ and for $a=0.60 $ GeV$^3$ corresponding to $\langle 0|\bar q q|0\rangle=(-248 $MeV$)^3$. They are shown in Table I for the case when only the terms $A_0$ and $B_3$ are included, and in Table II, where the contribution $A_4$ is added.
Here we employ the numerical value $b=0.5$ GeV$^4$.

Since the contribution of $A_4$ is numerically small, we do not vary the value of $b$.
We see that the nucleon mass becomes closer to the physical value when we include the condensate with dimension $d=4$. It also becomes closer to the physical value while we include the radiative corrections.
The accuracy of the solution  for the case when radiative corrections are included perturbatively is illustrated by Fig.~1.

In. Fig.~2\,$a$-$c$ we show dependence of the parameters $m$, $\lambda^2$ and $W^2$ on the actual value of the scalar condensate for the same case with the largest dimension of the condensates involved $d_{max}=3,4$ and $d_{max}=9$
\cite{6}.

The QCD sum rules provide a solution with observed  value of the nucleon mass if the terms with the dimension up to $d=9$ are included \cite{3}
\begin{equation}
\Pi_q^{OPE}(q^2)=\sum_{n=0}^{n=8}A_n(q^2); \quad \Pi_I^{OPE}(q^2)=\sum_{n=3}^{n=9}B_n(q^2).
\label{26}
\end{equation}
Now we study the influence of the terms with $d>4$ by "smooth switching on" their contribution to the sum rules.
To do this we construct the operators
\begin{equation}
\Pi_q^{OPE}(q^2)=A_0+A_4+\beta(A_6+A_8); \quad \Pi_I^{OPE}(q^2)=B_3+\beta(B_6+B_9),
\label{26a}
\end{equation}
tracing dependence on the solutions on $0\leq\beta\leq 1$. The results are shown in Fig.~3 One can see that our solution
converts smoothly to a physical one at $\beta=1$. The solution with $\lambda=0$, $W^2=0$
turns to the unphysical one. The latter was studied in details in \cite{8}.

\section{Summary}

Taking into account only the condensates of the lowest dimension $d=3$ or $d=3,4$ one immediately finds a
trivial exact solution $\lambda^2=0$, $W^2=0$ for the nucleon QCD sum rules in framework of the "pole+continuum " model for the spectrum of the polarization operator. For a long time there was a general believe that this is the only solution  if the condensates with $d_{max}=4$ are included.

We find a
nontrivial solution for the QCD sum rules equations in this case. Assuming that $\lambda >0$ we find a set of values $m,\lambda^2$ and $W^2$, which minimize the function $\chi_N^2$ defined by Eq.(\ref{3x}).  It can be treated as the lowest order approximation to the OPE solution. This is important since the condensates of the lowest dimension $\langle 0|\bar q q|0\rangle$ and $\langle 0|\frac{\alpha_s}{\pi}G^{a\mu \nu}G^a_{\mu \nu}|0\rangle$ can be either calculated or related in terms of observables, while the condensates with larger dimensions are known with bigger uncertainties. The convergence of the OPE series is illustrated by Fig.~2. In Fig.~3 we show how the trivial solution turns to an  unphysical one during the smooth switching on of the condensates of higher dimension. In the same figure we show how our solution turns to the physical one.

The authors acknowledge support by the grant RFBR  12-02-00158.

{}

\clearpage

\newpage
\begin{table}
\caption{Solutions of the sum rules equations (\ref{22}) for $\Pi_q$ and (\ref{22a}) for $\Pi_I$ with the terms $A_0$ and $B_3$
on the LHS. For each value of $a$ the first line - radiative corrections are neglected (I); second line - radiative corrections are included in framework of the LLA (II); third line - the LLA terms and the nonlogarithmic corrections of the order $\alpha_s$ are taken into account (III); fourth line - all radiative corrections are included in the lowest order of $\alpha_s$ (IV).}

\begin{center}
\begin{tabular}
{|c|c|c|c|c|c|} \hline
$a$,\,GeV$^3$&&$m$,~MeV&$\lambda^2$,~GeV$^6$&$W^2$,~GeV$^2$&
$\chi^2_N$\\
\hline

&I&1.31&3.46&3.14&6.2(-2)\\

0.50 &II&1.44&5.15&4.56&9.2(-2)\\

&III&1.23&3.47&2.81&1.6(-2)\\
&IV&1.18&2.90&2.41&2.7(-2)\\

\hline
&I&1.36&4.35&3.47&8.1(-2)\\
0.55&II&1.49&6.49&5.27&1.4(-1)\\
&III&1.27&4.33&3.08&2.3(-2)\\
&IV&1.23&3.65&2.64&3.4(-2)\\
\hline

&I&1.40&5.37&3.84&1.0(-1)\\
0.60&II&1.53&8.01&6.19&2.0(-1)\\
&III&1.31&5.32&3.37&2.3(-2)\\
&IV&1.27&4.50&2.89&4.3(-2)\\
\hline

\end{tabular} \end{center}
\end{table}

\begin{table}
\caption{Solutions of the sum rules equations (\ref{22}) and (\ref{22a}) with the terms $A_0$ , $A_4$ and $B_3$
on the LHS. The structure is the same as in Table I.}

\begin{center}
\begin{tabular}
{|c|c|c|c|c|c|} \hline
$a$,\,GeV$^3$&&$m$,~MeV&$\lambda^2$,~GeV$^6$&$W^2$,~GeV$^2$&
$\chi^2_N$\\
\hline

&I&1.20&2.61&2.65&1.2(-2)\\
0.50 &II&1.35&4.12&3.94&2.2(-2)\\
&III&1.16&2.98&2.57&2.9(-2)\\
&IV&1.10&2.31&2.10&6.4(-3)\\
\hline
&I&1.25&3.34&2.96&1.9(-2)\\
0.55&II&1.40&5.26&4.51&4.1(-2)\\
&III&1.21&3.75&2.83&6.1(-3)\\
&IV&1.15&2.96&2.33&9.9(-3)\\
\hline

&I&1.30&4.18&3.29&2.9(-2)\\
0.60&II&1.45&6.56&5.21&7.0(-2)\\
&III&1.25&4.64&3.11&1.1(-2)\\
&IV&1.19&3.71&2.57&1.1(-2)\\
\hline

\end{tabular} \end{center}
\end{table}

\clearpage

\newpage

\section{Figure captions}

\noindent
Fig.~1. Consistency of the LHS and RHS of the sum rules for $a=0.55$ GeV$^3$. The terms $A_0$ and $A_4$ are included on the LHS of Eqs.~(\ref{22}) for $\Pi_q$, and the term $B_3$ is included on the RHS of (\ref{22a}) for $\Pi_I$. Radiative corrections of the order $\alpha_s$ are included perturbatively (IV).
Solid lines and dashed lines show the LHS-to-RHS ratio for Eqs.~(\ref{22}) and (\ref{22a}) correspondingly.\\

\noindent
Fig.~2. Dependence of the values of nucleon parameters on the value of quark scalar condensate for various values of the largest dimension $d_{max}$ of the condensates. The dotted, dashed and solid lines are for $d_{max}=3$, $d_{max}=4$ and  $d_{max}=9$ correspondingly. Radiative corrections of the order $\alpha_s$ are included perturbatively (IV)\\

\noindent
Fig.~3. Evaluation of solutions during the "smooth switching on" of the condensates of higher dimension.
Dashed line -- evaluation of the trivial solution to the unphysical one.
Dashed-dotted line -- evaluation of our solution to the physical one.
The radiative corrections of the order $\alpha_s$ are included perturbatively (IV), $a=0.55$ GeV$^3$.

\newpage

\begin{figure}
\centerline{\epsfig{file=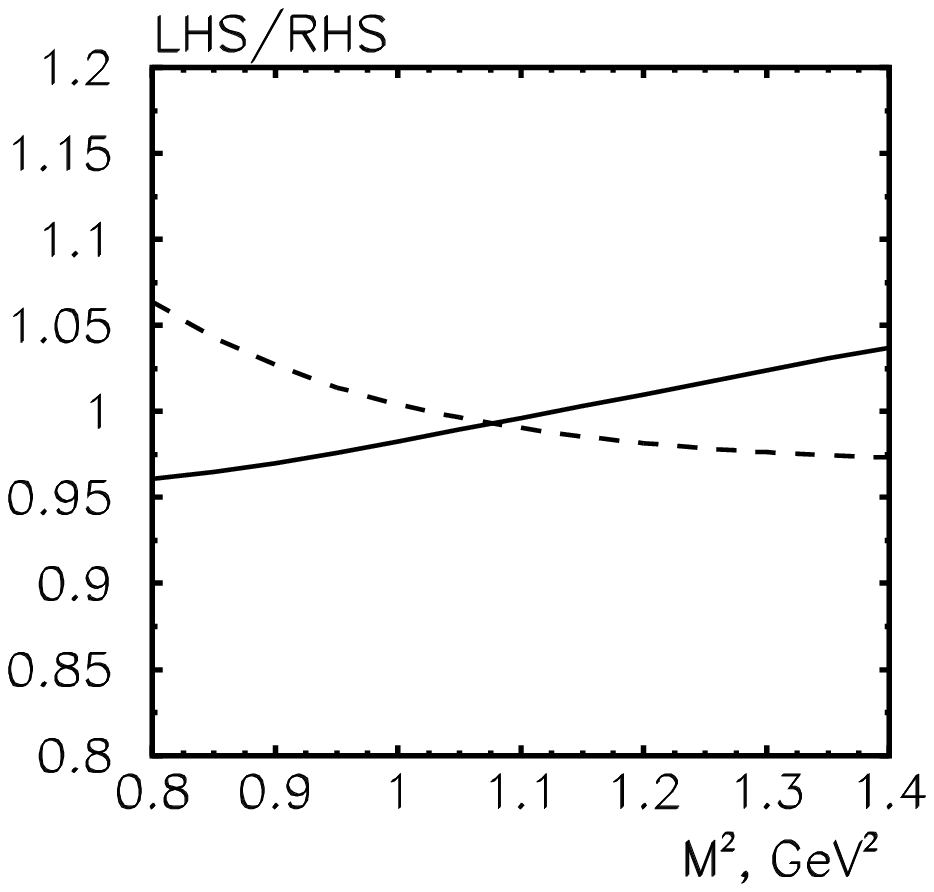,width=7cm}}
\caption{}
\end{figure}

\begin{figure}
\centerline{\epsfig{file=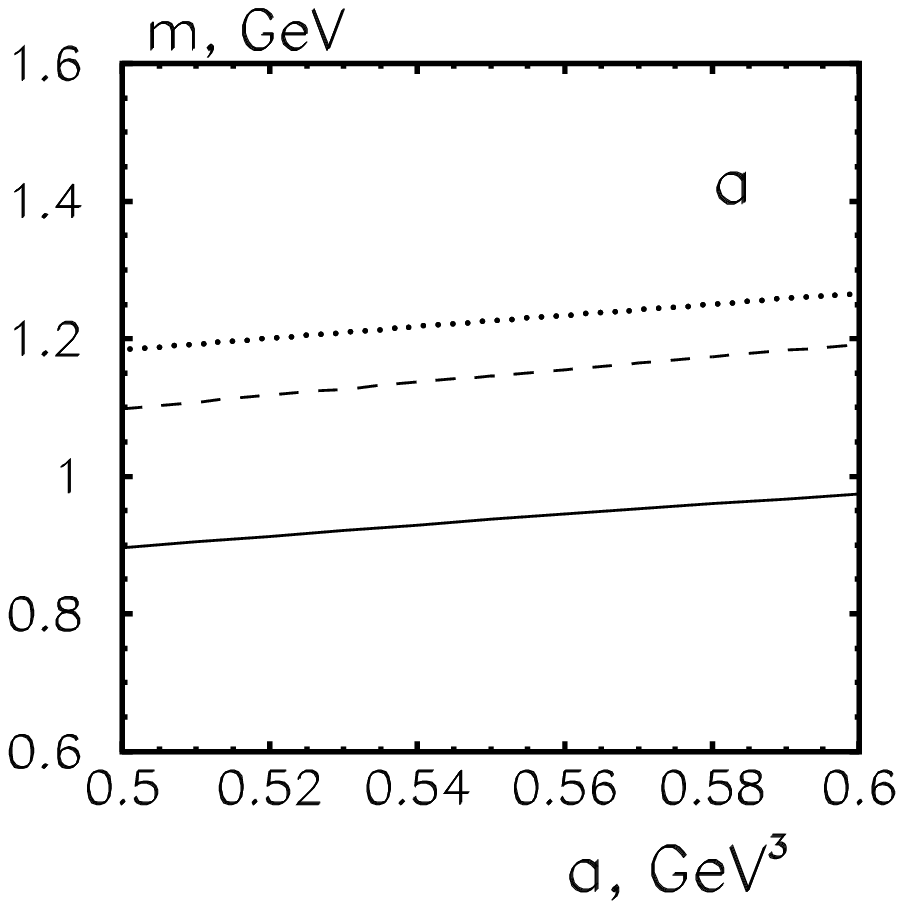,width=7cm}\epsfig{file=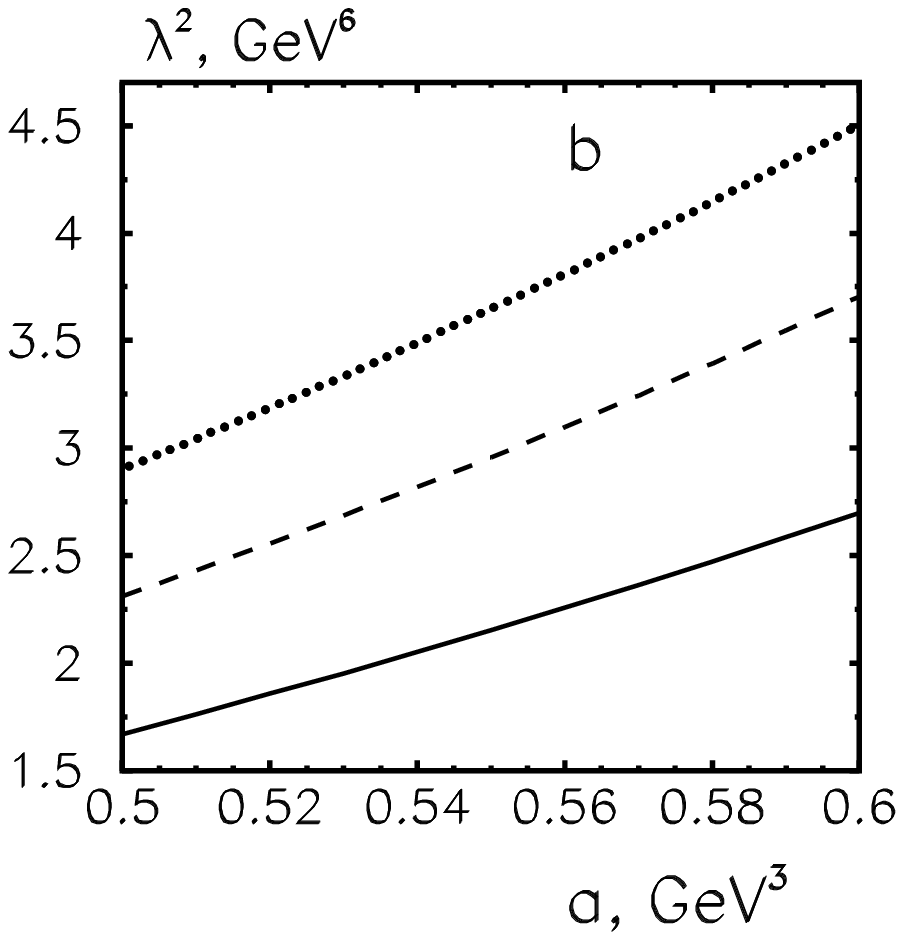,width=7cm}}
\vspace{1cm}
\centerline{\epsfig{file=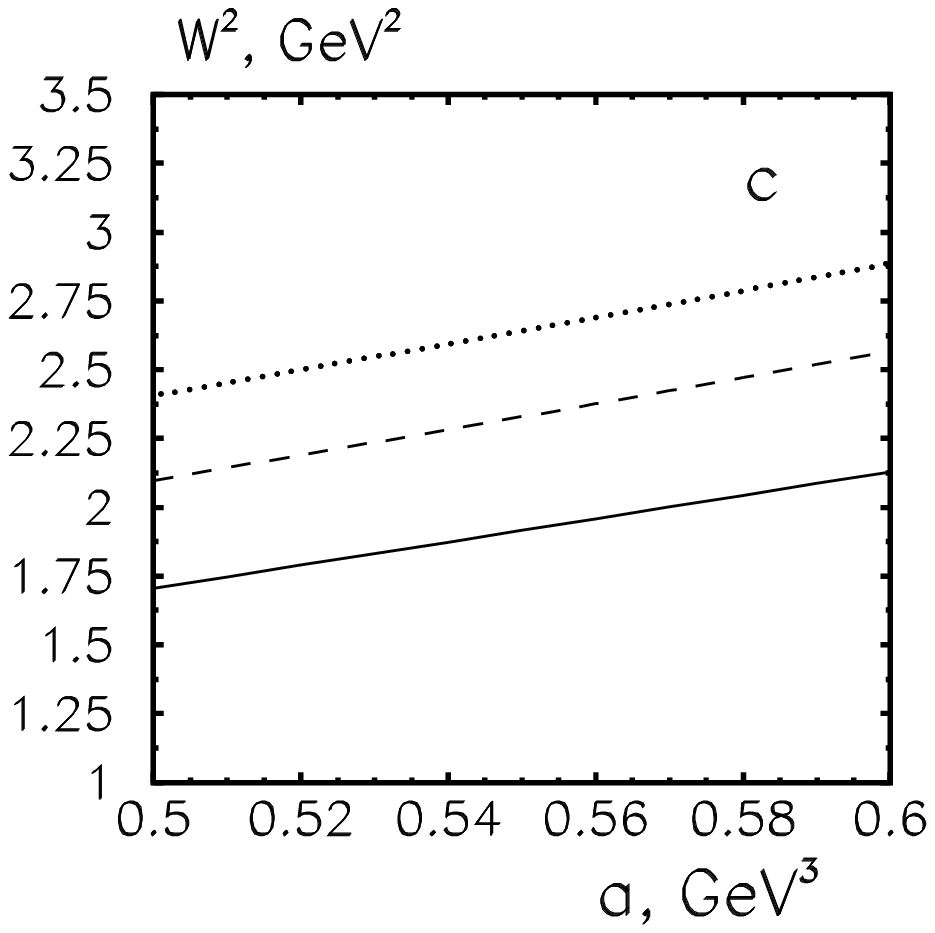,width=7cm}}
\caption{}
\end{figure}

\begin{figure}
\centerline{\epsfig{file=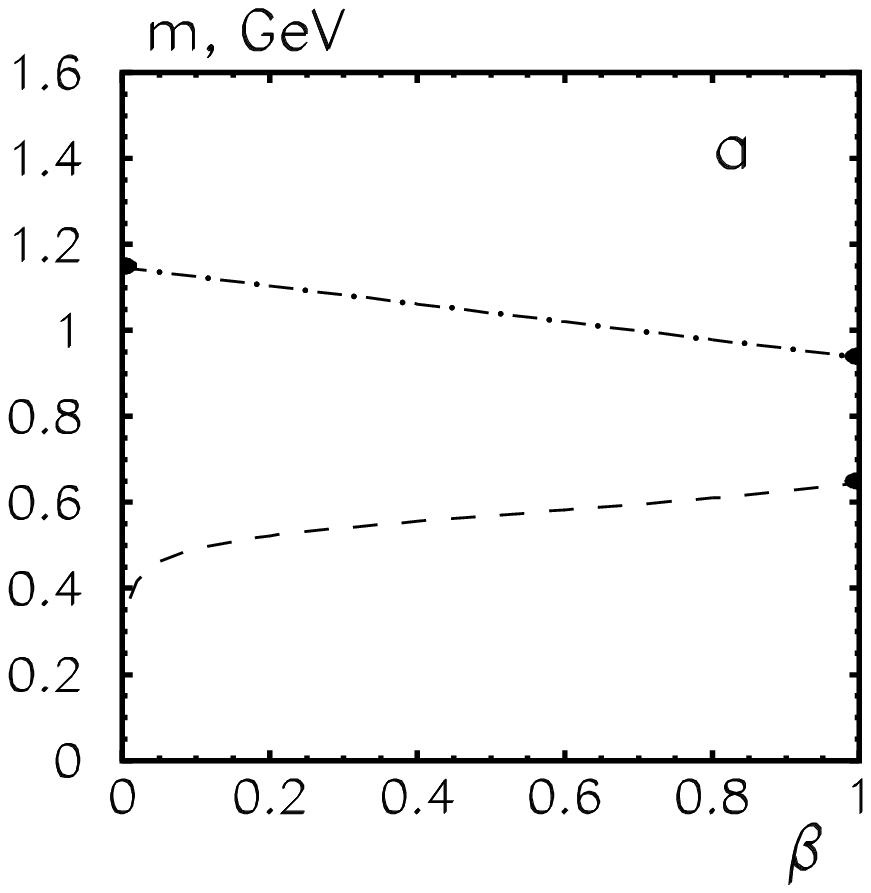,width=7cm}\epsfig{file=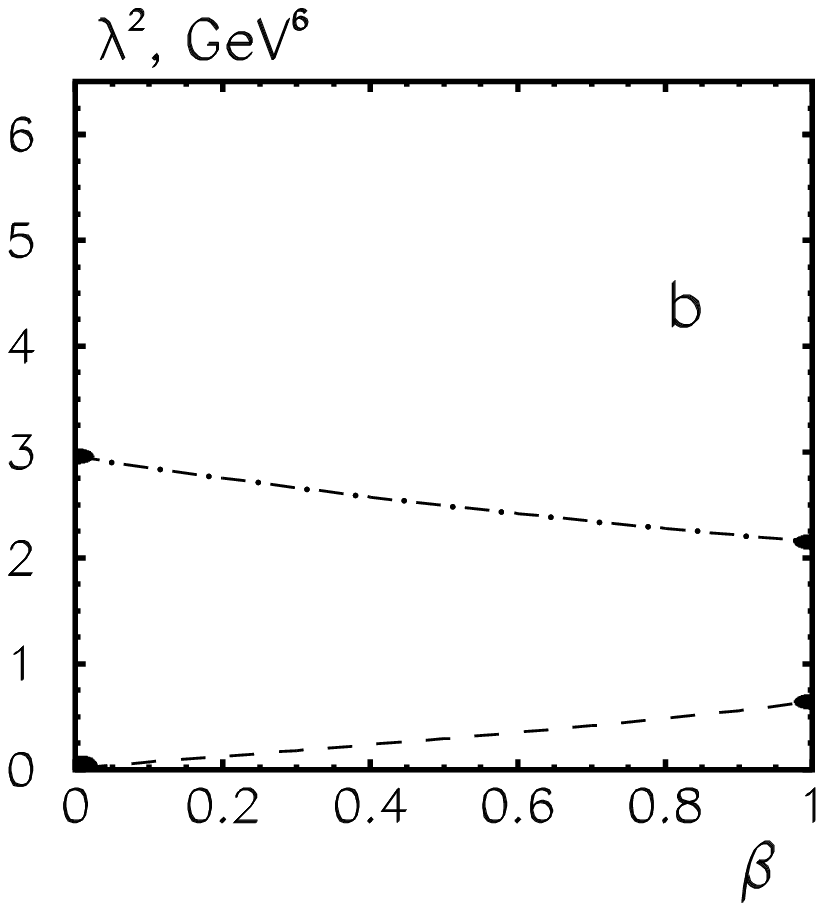,width=7cm}}
\vspace{1cm}
\centerline{\epsfig{file=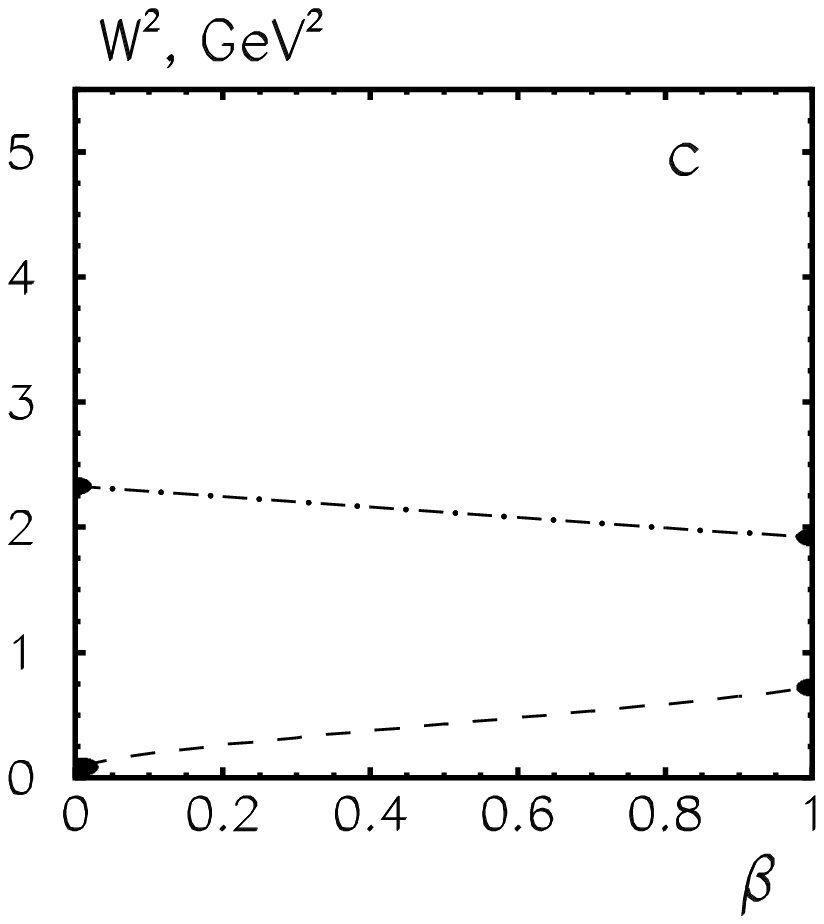,width=7cm}}
\caption{}
\end{figure}


\clearpage

\newpage

\begin{thebibliography}{}

\bibitem{1}M.A. Shifman, A.I. Vainshtein and V.I.~Zakharov, Nucl.
Phys. ~{\bf B147}, 385; 448; 519 (1979).

\bibitem{2} B.L. Ioffe, Nucl. Phys. ~{\bf B188}, 317 (1981); {\bf B191}, 591(E) (1981).

\bibitem{3} B.L. Ioffe, L.N. Lipatov and V.S.~Fadin, {\em Quantum
Chromodynamics} (Cambridge Univ. Press, 2010).

\bibitem{8} E.~G.~Drukarev, M.~G.~Ryskin, and V.~A.~Sadovnikova, Phys.
Rev. D~{\bf 80}, 014008 (2009).

\bibitem{4} D. V. Leinweber, Ann. Phys. {\bf 254}, 328 (1997).

\bibitem{5} B.L. Ioffe, Z. Phys. C {\bf18}, 67 (1983).

\bibitem{6} V.~A.~Sadovnikova, E.~G.~Drukarev and M.~G.~Ryskin, Phys.
Rev. D~{\bf 72}, 114015 (2005).

\bibitem{7} A. A. Ovchinnikov, A. A. Pivovarov and L. R. Surguladze, Int. J. Mod. Phys.
A {\bf 6}, 2025 (1991).



\end{thebibliography}
\end{document}